\documentclass[aps,prx,twocolumn,floatfix,superscriptaddress]{revtex4-1}  
\usepackage{graphicx}  
\usepackage{bm}        
\usepackage{amssymb}   
\usepackage{amsmath}
\usepackage{hyperref}
\usepackage[latin1]{inputenc}

 \usepackage{dcolumn}
 \usepackage{amsmath,amssymb,amsfonts}
 \usepackage{bm}
 \usepackage{latexsym}
 \usepackage{epstopdf}

\usepackage{color}

\newcommand{\bra}[1]{\left\langle #1\right|}
\newcommand{\ket}[1]{\left| #1\right\rangle}

\newcommand\kk{\mathbf{k}}
\newcommand\rr{\mathbf{r}}

\begin{document}
\title{Bound states in boson impurity models}
   \author{Tao Shi}
   \affiliation{Max-Planck-Institut f\"{u}r Quantenoptik Hans-Kopfermann-Str. 1, 
  85748 Garching, Germany }

   \author{Ying-Hai Wu}
   \affiliation{Max-Planck-Institut f\"{u}r Quantenoptik Hans-Kopfermann-Str. 1,
  85748 Garching, Germany }
  
   \author{A. Gonz\'{a}lez-Tudela}
   \affiliation{Max-Planck-Institut f\"{u}r Quantenoptik Hans-Kopfermann-Str. 1,
  85748 Garching, Germany }
 
\author{J. I. Cirac}
  \affiliation{Max-Planck-Institut f\"{u}r Quantenoptik Hans-Kopfermann-Str. 1,
 85748 Garching, Germany }

\date{\today}

\begin{abstract}
The formation of bound states involving multiple particles underlies many interesting quantum physical phenomena, such as Efimov physics or superconductivity. In this work we show the existence of an infinite number of such states for some boson impurity models. They describe free bosons coupled to an impurity and include some of the most representative models in quantum optics. We also propose a family of wavefunctions to describe the bound states and verify that it accurately characterizes all parameter regimes by comparing its predictions with exact numerical calculations for a one-dimensional tight-binding Hamiltonian. For that model, we also analyze the nature of the bound states by studying the scaling relations of physical quantities such as the ground state energy and localization length, and find a non-analytical behavior as a function of the coupling strength. Finally, we discuss how to test our theoretical predictions in experimental platforms such as photonic crystal structures and cold atoms in optical 
lattices.
\end{abstract}

\maketitle

\section{Introduction\label{sec:intro}}

Boson impurity models, where a two-level system (the impurity) is coupled to a bosonic bath, appear in the description of a variety of physical systems. In particular, they constitute a central paradigm in the field of quantum optics, where the impurity is an emitter (eg, an atom), and the bosonic bath corresponds to the modes of the electromagnetic field. Despite their simplicity, boson impurity models display a variety of basic phenomena. One of the most intriguing one is the existence of a single--excitation bound states (sEBS) for optical emitters interacting with photonic bandgap reservoirs \cite{john90a}, giving rise to interesting phenomena such as fractional decays or localization phase transitions. In this work we analyze the existence of many multi-excitation bound states (mEBS) in very generic boson impurity models, including those considered by John and Wang \cite{john90a} and other central problems in quantum optics. The existence of bound states lie at the heart many exotic phenomena like three-
body Efimov states \cite{efimov70a} but also very practical ones like Cooper pairs in superconductivity \cite{bardeen57a} or polarons \cite{holstein59a,holstein59b} in electron transport. Furthermore, the interest in mEBS is also triggered by the experimental progress in atom-nanophotonics integration \cite{goban13a,yu14a,lodahl13a,laucht12a,arcari14a,thompson13a},
as well as on the dramatic consequences played by sEBS on the generation of long-range interactions between atoms \cite{shahmoon13a,douglas15a,gonzaleztudela15b}.

In this work we will concentrate on a set of boson impurity models described by a Hamiltonian of the form
 \begin{equation}
 \label{Ham1}
 H = \Delta |e\rangle\langle e| + \sum_{\kk} \varepsilon_\kk a^\dagger_\kk a_\kk + \Omega \sum_\kk g_\kk (a_\kk^\dagger |g\rangle\langle e| + \mathrm{h.c.})
 \end{equation}
Here, $g,e$ represent the two-levels, the $a_\kk$ are annihilation operators of the bosonic bath, with a well-defined momentum $\kk$, and $g_\kk$ are the coupling constants, which are normalized to one, and that are typically very smooth functions of $\kk$. The most relevant parameters are the detuning $\Delta$, the coupling strength, $\Omega\ge 0$, and the dispersion relation, $\varepsilon_\kk$, which we will assume to correspond to a single band with bandwidth $W$.
In the quantum optical context, $\Delta$ describes the detuning between the two-level transition and the lowest energy of the bath Hamiltonian, and $\Omega$ the coupling strength in the rotating wave approximation limit. The third term thus describes the process in which the emitter is de-excited by emitting a photon into the bath.
Hamiltonian (\ref{Ham1}) models a number of relevant problems, ranging from spontaneous emission of a free atom, to the coupling of an emitter to a photonic crystal in any spatial dimension, as well as a single emitter in a high-Q cavity (Jaynes-Cummings model), which is recovered in the limit of $W\rightarrow 0$.

\begin{figure}[tb]
    \centering
    \includegraphics[width=0.44\textwidth]{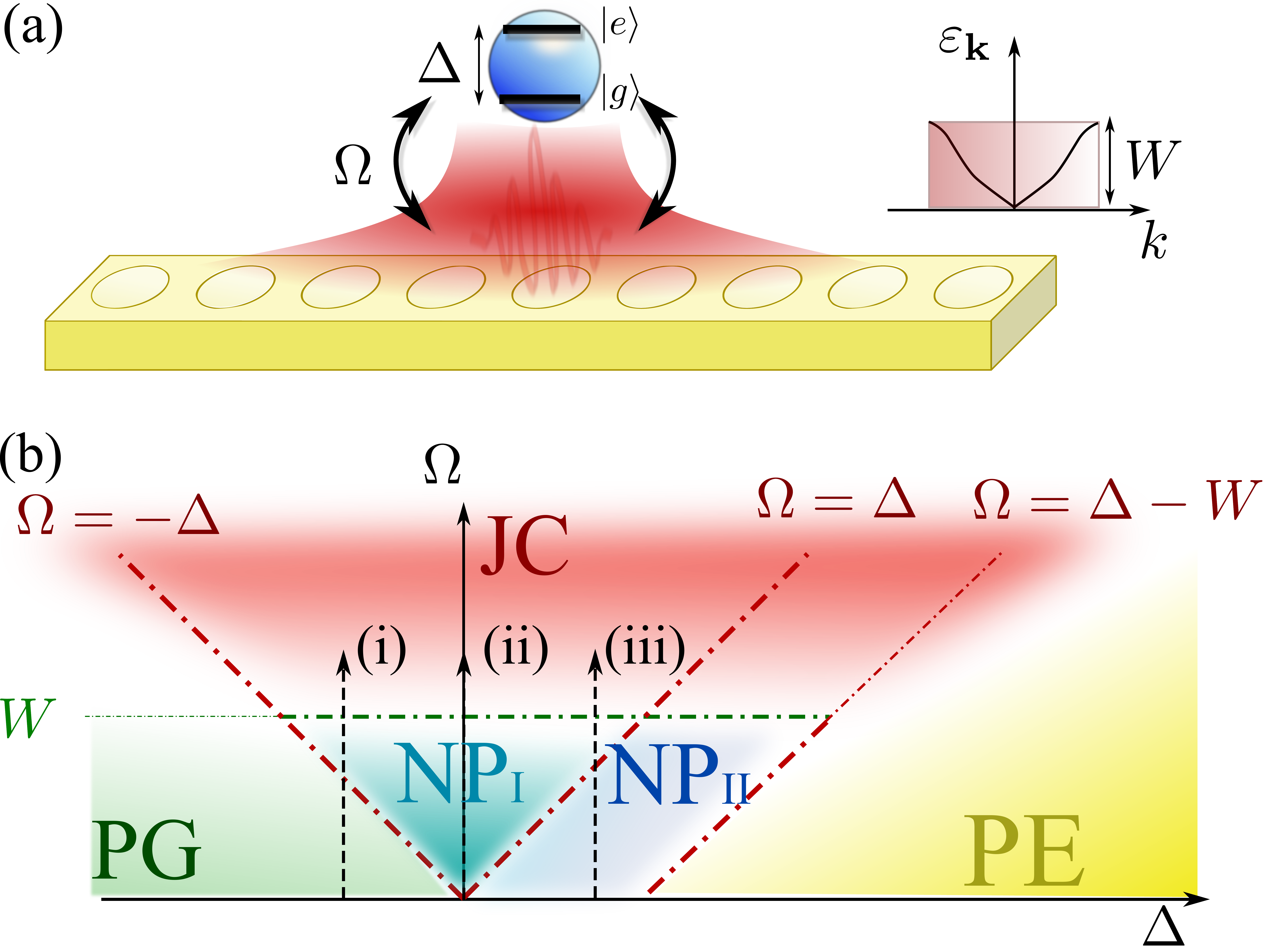}
    \caption{(a) Scheme of a single impurity with energy $\Delta$ coupled through $\Omega$ to a bath with dispersion relationship $\varepsilon_\kk$, a bandwidth $W$ with a localized bound state around it. (b) Parameter space of the one-dimensional $\cos(k)$ spectrum model as a function of $\Omega$ and $\Delta$. Different colors denote regions with different EBS origin: Perturbative Ground (PG, green), Perturbative Excited (PE,yellow), Non-Perturbative (NP, blue) and Jaynes-Cummings (JC,red), whereas the different tones (light/dark) represent different scaling regimes within the same origin:  NP$_{\mathrm{I}}$ and  NP$_{\mathrm{II}}$. See the text for further details. The three dashed arrows (i-iii) denote the paths that we use in the Figs.~\ref{fig2} and \ref{fig3}. }
    \label{fig1}
\end{figure}

More than twenty years ago John and Wang discovered that for certain dispersion relations \cite{john90a}, this Hamiltonian displays an exact eigenstate of the form
 \begin{equation}
 |B_1\rangle = \alpha |e\rangle |0\rangle + \beta |g\rangle |1\rangle,
 \end{equation}
where $|0\rangle,|1\rangle$ are states with zero and one bosons in the bath, respectively. In particular, the boson density vanishes far away from the position of the impurity, so that one can interpret that the boson is trapped by the impurity. The length at which this happens, {\em the localization length}, depends on the parameters of the model. In recent years, bound states with two excitations have been predicted for some particular dispersion relations \cite{zhou08a,shi09a,shi09b,kocabacs15a,schneider15a}. Furthermore, there is strong numerical evidence that analogous models \cite{sanchezburillo14a} (where the number of excitations is not conserved), may also possess bound states involving multiple excitations. Thus, some questions that naturally arise are if indeed, Hamiltonian (\ref{Ham1}) possesses multiple excitation bound states (mEBS), and if so, what is their origin, how to describe them precisely, and how to observe them experimentally.

In this paper we address all those questions. First of all, we show that for very generic dispersion relations, the boson impurity model (\ref{Ham1}) may support infinitely many bound states corresponding to different numbers of excitations. We investigate the origin of those bound states in the limit $|\Delta|\gg W,\Omega$, where one can view the impurity as creating a potential for the bosonic bath, in which the excitations may Bose-Einstein condense. We also postulate a three parameter family of (approximate) wavefunctions for the bound states in the form
 \begin{equation}
 |B_N\rangle = (\alpha |e\rangle\langle g| + \beta C^\dagger) A^{\dagger N-1} |g\rangle|0\rangle,
 \end{equation}
where $A$ and $C$ are linear combinations of the $a_\kk$ which depend on the dispersion relations. We confirm that those wavefunctions provide a very accurate description of the mEBS for a specific model in one spatial dimension by comparing their physical properties with the results obtained using advanced density-matrix renormalization group (DMRG) techniques \cite{white92a,schollwock11a}. In this model, we investigate the properties of the mEBS for all parameter regimes, and discover a region that cannot be described perturbatively, in which the energy and correlation length of the bound states are non-analytical functions of $\Omega$. Additionally, we  give exact expressions for the bound states for up to three excitations in the general case. Finally, we propose two different setups where the bound states could be prepared and observed. The first one uses atoms in optical lattices, where the role of the impurity is played by the absence/presence of an atom in an internal state, and the bath by the atoms in another internal state. The dispersion relation can be designed by
choosing the geometry of the lattice, and the value of the two other parameters, $\Delta$ and $\Omega$ can be easily tuned by changing the laser frequency and intensity. The second one corresponds to the scenario of an atom coupled to a photonic crystal and it will be much harder to observe. However, in view of the rapid experimental progress in different fronts \cite{goban13a,yu14a,lodahl13a,laucht12a,arcari14a,thompson13a}, it is not unforeseeble that some of those states or their consequences could be also observed in the near future.

The paper is structured as follows: In Section \ref{sec:model} we introduce the general model and requirements that we use along the manuscript. In Section \ref{sec:conditions} we derive the conditions for the existence of the mEBSs and introduce a variational ansatz to describe them. Then, in Section \ref{sec:onedexample} we study a particular example of a one-dimensional bath with $\cos(k)$ dispersion relationship and do a thorough study by using both the variational ansatz and numerical calculations. We explore the parameter space and reveal the existence of different regimes, yielding different scaling of the energies and localization lengths of the mEBS with the relevant parameters of the system.  Finally, in Section \ref{sec:implementations} we show how to prepare and detect these mEBSs in two different implementations and in Section \ref{sec:conclusions} we conclude by summarizing the main results of the manuscript.

\section{Model\label{sec:model}}

The very generic model that will be used along the manuscript is schematically depicted in Fig.~\ref{fig1}(a). It consists of three parts: a single impurity, a $d$-dimensional bath of free bosons, and the coupling between impurity and bath. The impurity has two levels $\{\ket{g},\ket{e}\}$ described by the Hamiltonian $H_{\rm imp}=\Delta\sigma_{ee}$ (the operator $\sigma_{\alpha\beta}$ is defined as $\ket{\alpha}\bra{\beta}$). The bath of volume $V=L^d$ with energy dispersion $\varepsilon_\kk$ is described by the Hamiltonian $H_{\rm bath}=\sum_{\kk}\varepsilon_\kk a^\dagger_\kk a_\kk$, where $a_\kk^\dagger (a_\kk)$ is the creation (annihilation) operators of bosons in the bath. Without loss of generality, we assume that the spectrum has a lower bound ${\rm min}|\varepsilon_\kk|=0$ but its bandwidth $W$ may be finite or infinite ({\em e.g.}, in the case of $\varepsilon_\kk\sim|\kk|^2$ spectrum). The couplig between impurity and bath, with a strength determined by $\Omega$, is encoded in the Hamiltonian
\begin{eqnarray}
H_{\rm int} &=& \frac{\Omega}{\sqrt{V}}\sum_\kk \eta_\kk (a^\dagger_\kk\sigma_{ge}+{\rm h.c.}) \nonumber \\
            &=& \Omega \sum_{j} \eta_j (a_j^\dagger \sigma_{ge}+{\rm h.c.}),
\label{eq:hint}
\end{eqnarray}
where $a_j=\sum_\kk a_\kk e^{i\kk \cdot \rr _j}/\sqrt{V}$ and $\eta_\kk =\sum_j \eta_j e^{-i \kk \cdot \rr_j}$ represents the mode function of the bosonic field (it is related to $g_\kk$ in Eq.~\ref{Ham1} via $g_\kk=\sqrt{V}\eta_\kk$). The total Hamiltonian of our system is $H=H_{\rm imp}+H_{\rm bath}+H_{\rm int}$.

It is worth emphasizing that this class of models are ubiquitous and appear in systems ranging from atom coupled to photonic crystals waveguides \cite{john90a,shahmoon13a,douglas15a,gonzaleztudela15b}, superconducting qubits coupled to microwave resonators \cite{mlynek14a}, and cold atoms in state dependent optical lattices \cite{devega08a,navarretebenlloch11a}, as we will see in Section \ref{sec:implementations} when we discuss the preparation and detection of the mEBSs. Among the different implementations, a one-dimensional bath with dispersion relation $\varepsilon_k =2J[1-\cos(k)]$ and $\eta_j=\delta_{j0}$ (this can be obtained in a tight-binding model with nearest neighbor hopping) is especially attractive due to recent developments in atom waveguide QED systems \cite{goban13a,yu14a,lodahl13a,laucht12a,arcari14a,thompson13a}. Its properties will be studied 
thoroughly using analytical, variational, and numerical methods in Section \ref{sec:onedexample}.

\section{Conditions for the existence of bound states\label{sec:conditions}}

In this Section we derive the conditions for the existence of bound states for arbitrary dispersion relations and spatial dimensions. The Hamiltonian $H$ commutes with the total excitation number operator $N=\sigma_{ee}+\sum_\kk a^\dagger_\kk a_\kk$ so we can study the subspaces with different $N$ separately. We first revisit the bound state of one boson \cite{john90a,douglas15a,gonzaleztudela15b} in Section \ref{sec:conditionssingle} in a way more general than what previous works did. For the more interesting cases with multiple bosons, we divide the parameter space into several regimes as depicted in Fig.~\ref{fig1} (b). The division is based on the methods that we use to understand them as well as the different scaling behaviors of the physical quantities. It will be proved in Section \ref{sec:conditionsmany} using analytical methods that bound states exist in certain regimes. These results motivate us to introduce a variational ansatz in Section~\ref{sec:variational} that may be used in all the 
regimes.

\subsection{Single excitation bound state \label{sec:conditionssingle}}

This case has been considered before in the literature, but mainly focusing on baths with quadratic ($\epsilon_\kk \sim |\kk|^2$) or tight-binding ($\epsilon_\kk\sim \sum_{i} \cos(\kk_i)$) dispersions in different spatial dimensions \cite{john90a,douglas15a,gonzaleztudela15b}. However, it is instructive to revisit this problem because we can find the conditions for the existence of bound states independent of the model. The wavefunction of one boson in the system can be written as
\begin{equation}
\ket{B_1} =\left(u_{e}\sigma_{eg}+\sum_{\kk}f_{\kk}a_{\kk}^{\dagger}\right) \ket{g}\ket{0}
\end{equation}
where $\ket{0}$ is the vacuum state of the bath. Bound states would appear if the secular equation $H\ket{B_1}=E_1\ket{B_1}$ has solutions that lie outside of the bath spectrum ({\em i.e.} $E_{1}<{\rm min}|\varepsilon_\kk|$ or $E_{1}>{\rm max}|\varepsilon_\kk|$. For concreteness, we focus on the low energy sEBS with $E_{1}<{\rm min}|\varepsilon_\kk|$ but the method presented below can also be used to derive the conditions for the upper ones.

By defining the function
\begin{equation}
F_{1}(E)=E-\Delta -\Omega ^{2}\int \frac{d^{d}\kk}{(2\pi )^{d}}\frac{\left\vert \eta _{\kk}\right\vert ^{2}}{E-\varepsilon _{\kk}}\,.
\label{eq:1bs}
\end{equation}
the existence condition of sEBS is that  $F_{1}(E)=0$ must have solutions that lie outside of the bath spectrum. One can show that $F_{1}(E)$ has two properties: i) it is a monotonically increasing function; ii) $F_{1}(E)\rightarrow -\infty $ as $E\rightarrow -\infty $. Thus, if we have $F_{1}(0)>0$, there must be only one solution to $F_{1}(E)=0$ and a unique bound state with $E_{1}<0$ appears. Therefore, the existence or absence of bound state is essentially determined by the integral
\begin{equation}
I_{0}=\int \frac{d^{d}\kk}{(2\pi )^{d}}\frac{\left\vert \eta_{\kk}\right\vert ^{2}}{\varepsilon _{\kk}}\,.
\end{equation}
For example, for one- and two-dimensional baths with tight-binding and quadratic dispersions $I_0\rightarrow \infty$ due to an infrared divergence (as long as $|\eta_\kk|^2$ do not vanishes at $|\kk|=0$ or  is a highly oscillatory function). In these cases, there always exists a sEBS irrespective of $\Delta$, $\Omega$ or $W$. However, if $I_0$ converges to a finite positive value, which happens for a three-dimensional bath with tight-binding or quadratic dispersions ($I_0=0.253$ in these cases), a bound state exists only if $F_{1}(0)=-\Delta+\Omega^{2}I_{0}>0$. 

\subsection{Multiple excitations bound states in strong and weak coupling regimes\label{sec:conditionsmany}}

The analysis presented above can be extended to systems with multiple excitations. The wavefunction in the subspace with $N$ excitations can be written as
\begin{equation}
  \label{eq:st}
\ket{B_N} =\ket{\Psi_{e}}\ket{e}+\ket{\Psi_{g}}\ket{g},
\end{equation}
where the two states $\ket{\Psi_{g}}$ and $\ket{\Psi_{e}}$ dress the impurity atom when it is in $\ket{g}$ and $\ket{e}$ respectively. The existence of bound states and their properties can be understood analytically in some of the regimes depicted Fig.~\ref{fig1}(b) where the couplings are very strong or very weak.

For example, when the bath-impurity coupling is very strong [the red regime in Fig.~\ref{fig1}(b) with $\Omega\gg |\Delta|,W$, the dominant contribution to the Hamiltonian is the coupling between the impurity and a collective bath mode $A_{\mathrm{JC}}^{\dagger}=\sum_{j}\eta _{j}a^{\dagger}_{j}$ so the physics is the same as in the well-known \emph{Jaynes-Cummings} model \cite{jaynes63a}
\begin{equation}
\ket{B_N}=\frac{1}{\sqrt{\mathcal{N}_{\rm JC}}} \left( |E_{N}| A^{\dagger{N-1}}_{\rm JC} \ket{e} -\sqrt{N} \Omega A^{\dagger^{N} }_{\rm JC} \ket{g} \right)
\end{equation}
where $\mathcal{N}_\mathrm{JC}$ is obtained by imposing the normalization of $\ket{B_N}$ and $E_N=(\Delta-\sqrt{\Delta ^{2}+4N\Omega^{2}})/2$. The other couplings to the bath are weak perturbations that enters to $E_N$ as a frequency shift of smaller order than the leading term. 

For the opposite cases where the impurity-bath coupling is weak [the darker green and yellow regimes in Fig.~\ref{fig1}(b) with $\Omega\ll|\Delta|$ and $\Omega\ll\Delta-W$, the coupling $H_{\rm int}$ is a small perturbation to both the impurity and bath Hamiltonians so we can eliminate it to the first order of $\Omega$ using a Fr\"{o}hlich transformation \cite{supmat}. This leads to the effective Hamiltonian
\begin{eqnarray}
\label{eq:effham}
&& H_{\rm eff} = H_{\rm imp} + \frac{\Omega^{2}}{V} \sum_{\kk}\frac{|\eta _{\kk}|^{2}\sigma_{ee}}{\Delta-\varepsilon_{\kk}} + H_{\rm bath} \nonumber \\
&& + \frac{\Omega^{2}}{2V} \sum_{\kk,\kk^{\prime}} \eta^{\ast}_{\kk^{\prime}} \eta_{\kk} \left( \frac{1}{\Delta-\varepsilon_{\kk^{\prime}}} + \frac{1}{\Delta-\varepsilon_{\kk}} \right) a^{\dagger}_{\kk} a_{\kk^{\prime}} \sigma_{z}
\end{eqnarray}
with $\sigma_z=(\sigma_{ee}-\sigma_{gg})/2$ and from which it emerges a very clear physical picture: on the one hand, $H_{\rm int}$ induces a frequency shift on the impurity energy (usually called Lamb shift), while,  more importantly, $H_{\rm int}$ creates also a non-local potential for the excitations that localizes them around the impurity giving rise to a mEBS.

We term the perturbative regime with negative (positive) detuning $\Delta$ [see the green (yellow) regime in Fig.~\ref{fig1}(b)] as {\em ground} ({\em excited}) because in this regime the $\ket{\Psi_{g}}\ket{g}$ ($\ket{\Psi_{e}}\ket{e}$) part in Eq.~(\ref{eq:st}) is very small. The effective Hamiltonian can be projected to the $\ket{e}$ ($\ket{g}$) state in the PG (PE) regime and results in $H^{e}_{\rm eff}$ ($H^{g}_{\rm eff}$). The bath component $\ket{\Psi_{e}}=\frac{A^{\dagger{N-1}}_{e}\ket{0}}{\sqrt{(N-1)!}}$ in the PG regime, whereas the bath component $\ket{\Psi_{g}}=\frac{A^{\dagger{N}}_{g}\ket{0}}{\sqrt{N!}}$ in the PE regime. $A^{\dagger}_{e}$ and $A^{\dagger}_{g}$ are collective bosonic bath operators whose explicit forms can be found from the secular equation $H^{s}_{\rm eff}A^{\dagger}_{s}\ket{0}=E^{s}_{1}A^{\dagger}_{s}\ket{0}$ ($s=e$ or $g$).

The bosons form a bound state if the secular equation have solutions with energy values $E^{s}_1<0$. As for the case with one excitation, the existence condition can be formulated using two functions $F^{s}_N(E)$ that are defined in a way similar to Eq.~(\ref{eq:1bs}) \cite{supmat}. We find that mEBS exist in the PG and PE regimes if 
\begin{eqnarray}
F^{s}_{N}(0) = \left[1-\frac{w_1(\Delta)}{|\Delta|}\right]^{2} -\left[2+\frac{w_2(\Delta)}{|\Delta|}\right]\frac{\Omega^{2}I_{0}}{2|\Delta|} < 0
\end{eqnarray}
where
\begin{equation}
w_\alpha(E)=\frac{\Omega ^{2}}{2}\int \frac{d^{d}\kk}{(2\pi )^{d}}\frac{\varepsilon_\kk^{\alpha-1}|
\eta _{\kk}|^{2}}{(E-\varepsilon _{\kk})^\alpha}\,.
\end{equation}

If $I_{0}\rightarrow+\infty$ due to infrared divergence, we have $F^{s}_{N}(0)\rightarrow -\infty$ and mEBS exist in both regimes. For the perturbative ground regime, the ground state energy is $E^{e}_N=E_{1}+(N-1)E^{e}_{1}$ and the wave function $\ket{B_N}$ is
\begin{equation}
\label{eqmain:NB1}
\frac{A^{\dagger{N-1}}_{e}\ket{0}\ket{e}}{\sqrt{(N-1)!}} + \frac{C^{\dagger}_{e}A^{\dagger^{N-1}}_{e}\ket{0}\ket{g}}{\sqrt{(N-1)!}}
\end{equation}
(see \cite{supmat} for the explicit form of the operator $C^{\dagger}_{e}$). For the PE regime, the ground state energy is $E^{g}_N=NE^{g}_1$ and the wave function $\ket{B_N}$ is
\begin{equation}
\label{eqmain:NB2}
\frac{D_{g} A^{\dagger{N-1}}_{g}\ket{0}\ket{e}}{\sqrt{(N-1)!}} - \frac{A^{\dagger{N}}_{g}\ket{0}\ket{g}}{\sqrt{N!}}
\end{equation}
(see \cite{supmat} for the explicit form of the number $D_{g}$).

The analysis presented above relies on perturbation theory, but the bound states survive even if we move to the $\Omega>|\Delta|$ regime for negative $\Delta$ and the $\Omega>\Delta$ and $\Delta<W$ regime for positive $\Delta$. This is because increasing the coupling strength $\Omega$ for all values of $\Delta$ and reducing $\Delta$ for positive $\Delta$ makes the bound states more localized. In summary, we conclude that the mEBS exist in all parameter regimes if $I_{0}\rightarrow +\infty$. 

For baths with tight-binding and quadratic dispersions, $I_{0}$ diverges in one and two dimensions but converges in three dimensions. This means that sEBS and mEBS exist in all regimes in one and two dimensions but they only emerge for $\Omega$ larger than a critical value (which depends on $\Delta$) in three dimensions. Moreover, because $F^s_N(E)$ and $F_1(E)$ have very different forms, we may have only sEBS but not mEBS in certain regimes in three dimensions. 

\subsection{Variational ansatz for multiple excitation bound states in all regimes \label{sec:variational}}

In the previous Section, we show how for certain regimes defined by very strong or very weak couplings the existence of the mEBS can be guaranteed analytically. However, it is obviously desirable to have a way of describing the whole parameter space of Fig.~\ref{fig1}(b), including what we denoted as non-perturbative regimes (in blue) where both the weak/strong coupling expansions fail. Inspired by the perturbative wavefunctions of Eqs. \ref{eqmain:NB1} and \ref{eqmain:NB2}, we introduce the following variational ansatz for the ground and excited bosonic components of $\ket{B_N}$ as:
\begin{eqnarray}
\label{eqmain:ansatz}
\ket{\Psi_{e}} &=& \alpha \frac{A^{\dagger{N-1}}}{\sqrt{(N-1)!}}\ket{0} ,\nonumber \\
\ket{\Psi_{g}} &=& \left[ \beta \frac{A^{\dagger{N}}}{\sqrt{N!}} + \gamma \frac{A^{\dagger{N-1}}B^{\dagger}}{\sqrt{(N-1)!}} \right] \ket{0},
\end{eqnarray}
where $v=(\alpha,\beta,\gamma)^T$ is a vector of variational parameters that allows us to interpolate between the different parameter regimes and  $A^\dagger,B^\dagger=\sum_\kk \varphi_{A,B}(\kk)a^\dagger_\kk$ are two orthogonal collective bosonic modes of the bath. In order to obtain the wavefunction and $v$, one must minimize the energy of the variational ansatz $E_{\mathrm{GS}}=\bra{B_N}H\ket{B_N}$ under the constraint $v^{T}{\cdot}v=1$ and $[A,A^{\dagger}]=[B,B^{\dagger}]=1$ and $[A,B^{\dagger}]=0$. This procedure leads to the following coupled Gross-Pitaevski type equation for $\varphi_{A,B}(\kk)$:
\begin{equation}
\mathcal{H}_{0}
\left(
\begin{array}{c}
\varphi_{A}(\mathbf{k}) \\ 
\varphi_{B}(\mathbf{k})
\end{array}
\right) + \frac{\Omega\eta _{\mathbf{k}}}{\sqrt{V}}\alpha
\left( 
\begin{array}{c}
\sqrt{N} \beta \\ 
\gamma
\end{array}
\right) =\mu 
\left(
\begin{array}{c}
\varphi _{A}(\mathbf{k}) \\ 
\varphi _{B}(\mathbf{k})
\end{array}
\right),
\label{eq:GP2}
\end{equation}
where $\mu$ and $\mathcal{H}_0$ are defined in~\cite{supmat}. Remarkably, the structure of these coupled equations already determines that the wavefunctions are the superposition of two functions $\eta _{\mathbf{%
k}}/(e_{1,2}-\varepsilon _{\mathbf{k}})$ as:
\begin{equation}
\varphi _{M=A,B}(\mathbf{k})=\frac{1}{\sqrt{V}}\sum_{\mu=1,2}\frac{c_{M,\mu}}{\sqrt{\mathcal{N}_{\mu\mu}}}\frac{\eta _{\mathbf{k}}}{e_{\mu}-\varepsilon
_{\mathbf{k}}},
\label{eq:phs}
\end{equation}
where $c_{M,2}=t_{M}c_{M,1}$. The expressions of the coefficients $c_{M,\mu}$ and the normalizations $\mathcal{N}_{\mu\mu}$ are given in full detail in ~\cite{supmat}. With this insight, we minimize the energy $E_{\rm GS}$ with respect to $\{t_A,e_1,e_2\}$ to obtain the bound state energy $E_N={\rm min} E_{{\rm GS}}(t_A,e_1,e_2)$, the vector $v$, and the corresponding wavefunction~\cite{supmat}.

\section{One-dimensional tight-binding bath: variational and exact results \label{sec:onedexample}}

\begin{figure}[tb]
	\centering
	\includegraphics[width=0.4\textwidth]{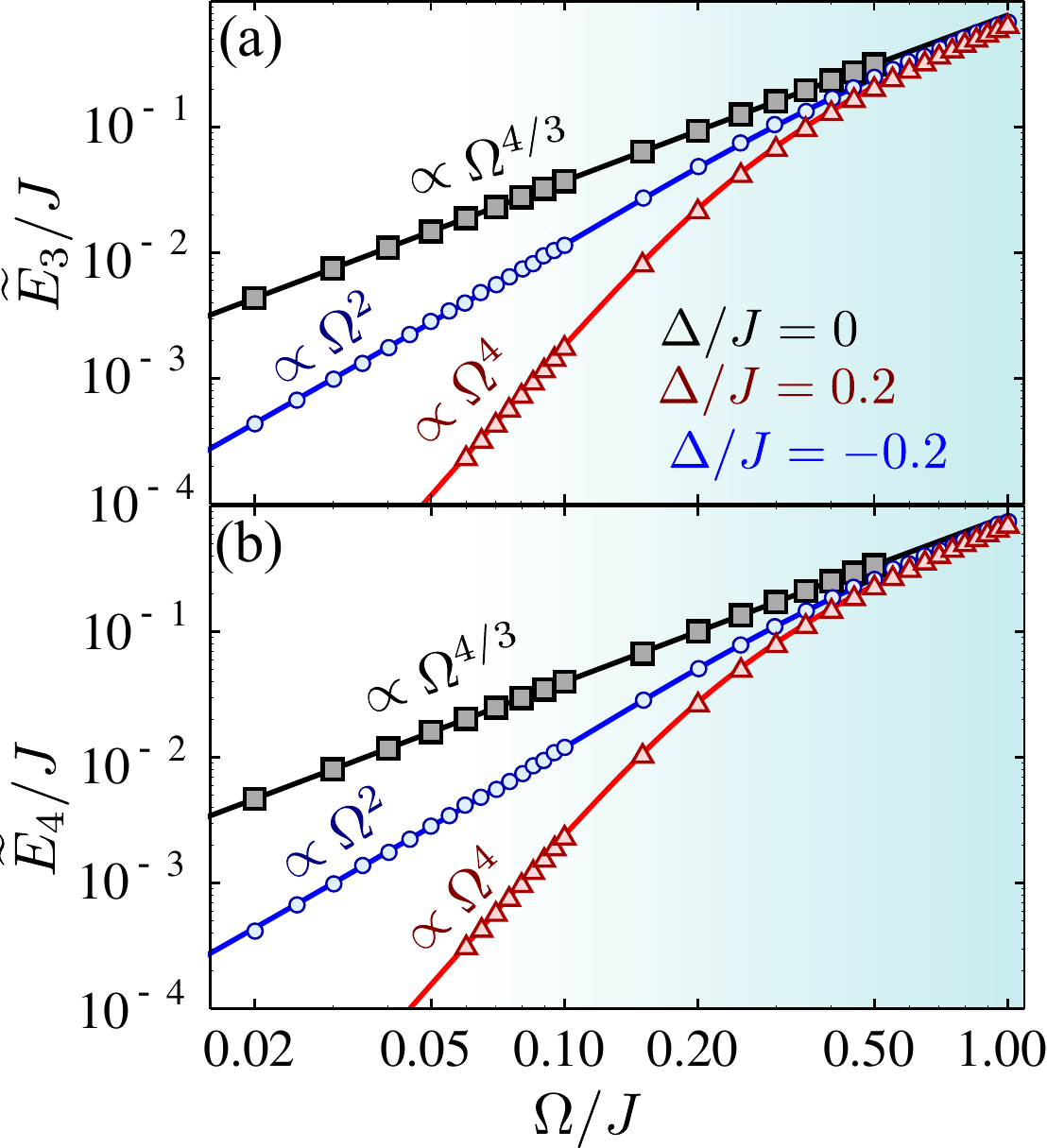}
	\caption{(a-b) Comparison of the bound state energies $\widetilde{E}_N$ given by the variational ansatz (solid lines) with infinite bath size and the DMRG results (markers) in the $N=3$ [Panel (a)] and $N=4$ [Panel (b)] subspaces. The three different curves correspond to the three different paths depicted in Fig.\ref{fig1}, i.e., choosing a detuning $\Delta/J=-0.2$, $\Delta/J=0$ and $\Delta/J=0.2$.}
	\label{fig2}
\end{figure}

\begin{figure}[tb]
	\centering
	\includegraphics[width=0.4\textwidth]{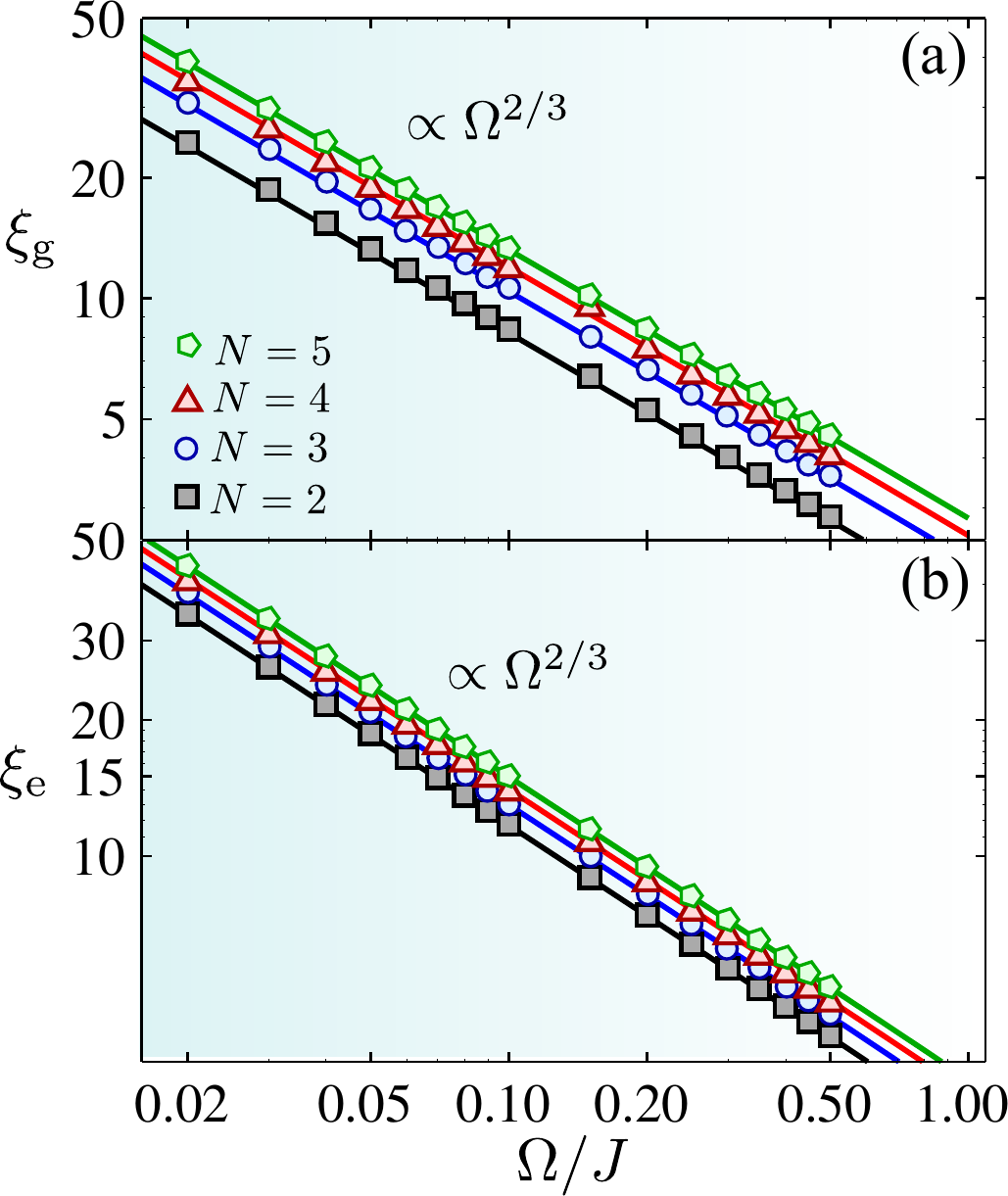}
	\caption{(a-b) Comparison of the localization lengths $\xi_{g}$ (a) and $\xi_{e}$ (b) given by the variational ansatz (solid lines) with infinite bath size and the DMRG results (markers) with $1000$ bath sites at $\Delta/J=0$ in the $N=2,3,4,5$ subspaces.}
	\label{fig3}
\end{figure}

In this Section, we study a particularly relevant model for which the bath is a one-dimensional lattice described by the tight-binding Hamiltonian $H_{\rm bath}=-J\sum_{\langle{jm}\rangle} (a^{\dagger}_j a_m + {\rm h.c.})$ (where $\langle{jm}\rangle$ denotes nearest neighbors) and the coupling between impurity and bath only occurs on one site, which we call the zeroth site, i.e., $\eta_j=\delta_{j\,0}$.We have solved this model exactly in the $N=2,3$ subspace~\cite{supmat} and using the DMRG algorithm~\cite{white92a,schollwock11a} to compute the ground states in the $N=2,3,4,5$ subspaces. The exact solutions can be obtained for periodic boundary conditions but the DMRG calculations are performed for open boundary conditions as they are more suitable for the algorithm. The boundary effect is negligible when the number of lattice sites is large enough (we have studied systems with up to 1000 bath sites). The $A^\dagger$ and $B^\dagger$ modes in Eq.~(\ref{eqmain:ansatz}) can 
be expanded using the bath operators as $\sum_{j}\varphi_{M,j}a^{\dagger}_{j}$ ($M=A$ or $B$), where the coefficients are
\begin{equation}
\varphi_{M,j} = \sum_{\mu=1,2} c_{M,\mu} \sqrt{\frac{1-x^{2}_{\mu}}{1+x^{2}_{\mu}}} x^{|j|}_{\mu}
\end{equation}
with $x_{\mu}+x^{-1}_{\mu}=2-e_{\mu}$. To confirm the validity of our variational ansatz, we compare some physical quantities given by analytical, numerical, and variational approaches. In addition to the ground state energy, we also study the experimentally measurable localization lengths of the bath components defined as ($s=e$ or $g$)
\begin{equation}
\xi_{s} = \sqrt{\frac{\sum_{j} j^{2} \bra{\Psi_{s}} a^{\dagger}_{j} a_{j} \ket{\Psi_{s}}} {\sum_{j}\bra{\Psi_{s}} a^{\dagger}_{j} a_{j} \ket{\Psi_{s}}}}.
\end{equation}
We choose three different paths in the parameter space labeled as (i-iii) in Fig.~\ref{fig1}(b), which cover a wide range of $\Omega$ at three different detunings $\Delta$. In Fig.~\ref{fig2}, we compare the modified ground state energies $\widetilde{E}_N=|E_N- H(-\Delta)\Delta|$ [where $H(x)$ is the step function] given by variational (solid lines) and DMRG (markers) calculations for the $N=3,4$ cases. In Fig.~\ref{fig3}, we compare the localization lengths given by variational (solid lines) and DMRG (markers) calculations for the $N=2,3,4,5$ cases. The perfect agreement in all cases clearly demonstrates the power of our variational ansatz.

We can provide a more detailed characterization of Fig. \ref{fig1}(b) by establishing scaling relations of physical quantities. This can be done using variational ansatz and DMRG results which leads to the scaling relations presented in Table~\ref{tab:scaling}. However, we should be careful when trying to extract the scaling exponents as they are only well defined in certain limits. Therefore, Fig ~\ref{fig1}(b) must not be understood as a phase diagram but rather as an indication of the different scaling behaviors of the system, which are only well defined in the corresponding limits.

To illustrate the different behaviors, we focus the discussion on the three paths depicted in Fig.~\ref{fig1}(b) as these represent the experimentally most relevant regimes that occur when $\Omega\ll W$. For path (i) with $\Delta/J=-0.2$, the system changes from the PG regime with $\widetilde{E}_N\propto\Omega^2$ to the  NP$_{\mathrm{I}}$ regime. This change is also manifested in the variational parameters: we have $1-|\alpha|\propto \Omega^2$ in the PG regime (so most of the weight in the wavefunction comes from $\ket{\Psi_{e}}$) but $|\alpha|,|\beta|,|\gamma|$ of the same order $\sim O(1)$ in the  NP$_{\mathrm{I}}$ regime. For path (ii) with at $\Delta/J=0$, we only move within the  NP$_{\mathrm{I}}$ regime. For path (iii) with $\Delta/J=0.2$, the system changes from the NP$_{\mathrm{II}}$ regime with $\widetilde{E}_N{\propto}N\Omega^4$ to the NP$_{\mathrm{I}}$ regime. $\widetilde{E}_N$ in the NP$_{\mathrm{I}}$ regime exhibits non-analytical behavior because it scales as a fractional power of $\Omega$ in this regime. The exponent is only well defined if $\Omega\ll|\Delta|$ which is satisfied by path (ii) where $\widetilde{E}_N\propto \Omega^{4/3}$, but not by paths (i) and (iii). The mEBS in the NP$_{\mathrm{II}}$ regime can be understood as a Gutzwiller projected condensate. To see this, we note that $1-|\beta|\propto\Omega^4$ so the wavefunction basically describes a state with $N$ bosons in the $A$ mode. This state can be constructed by first putting all the bosons in the sEBS $\ket{B_1}$ and then projecting out the configurations with more than one boson in the $\ket{e}$ state.

\begin{table}[tb]
 \label{tab:scaling}
 \begin{tabular}{ |c|c|c|c| } 
    \hline
     & PG & NP$_{\mathrm{I}}$ &   NP$_{\mathrm{II}}$,PE   \\ 
    \hline
    $E_N$ &  $\Delta-\frac{\Omega^2}{\sqrt{4|\Delta|(|\Delta|+4J)}}$ & $\propto \Omega^{4/3}$ &  $\propto N \Omega^4$  \\ 
    \hline
    $1-x_1$ & $\frac{\Omega^2}{2\Delta J}$ & $\propto \Omega^{2/3}$ &   $\propto \Omega^2$  \\ 
    \hline
    $1-x_2$ & $\frac{\Delta}{2J}\left(1+\sqrt{1-4J^2/\Delta^2}\right)$ & $\propto \Omega^{2/3}$ &  - \\ 
    \hline
    $|\alpha|$ & $1-d_\gamma^2\Omega^2$ & $O(1)$ &  $d_\alpha \Omega^2$ \\ 
    \hline
     $|\beta|$ & $\propto \Omega^2$ & $O(1)$ &  $1-d_\alpha \Omega^4$ \\ 
     \hline
     $|\gamma|$ & $d_\gamma \Omega$ & $O(1)$ &  $\propto\Omega^4$ \\ 
     \hline
\end{tabular}
\caption{Scaling analysis of the relevant properties of EBS for the regimes with $\Omega\ll W$ depicted in Fig.~\ref{fig1}(b).}
\end{table}

\begin{figure}[tb]
	\centering
	\includegraphics[width=0.4\textwidth]{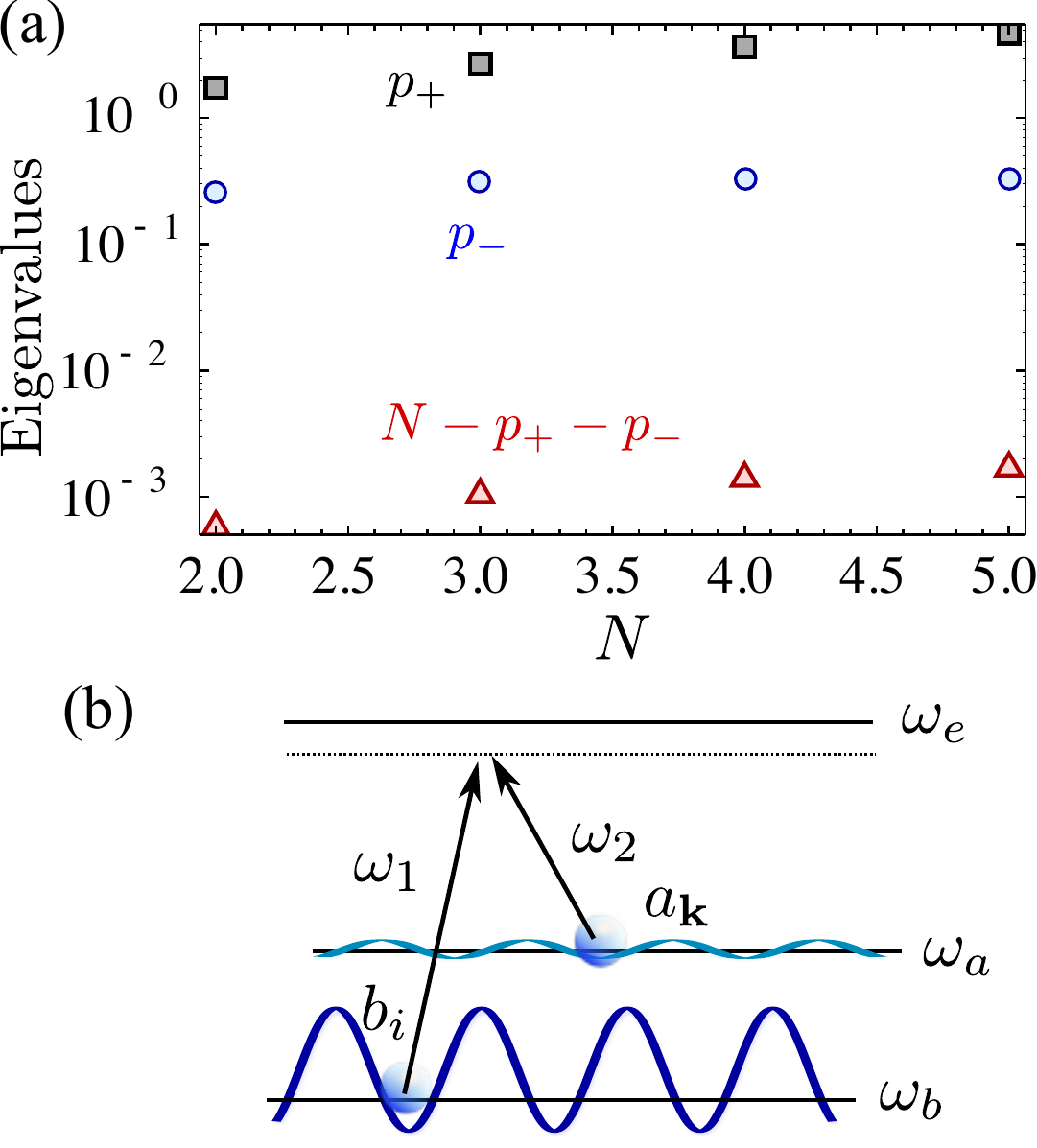}
	\caption{(a) Eigenvalues of the correlation matrix is shown for multi-EBS as a function of $N$ for $\Delta/J=0$ and $\Omega/J=0.5$. (b) Scheme of the implementation with state dependent optical lattices: two atomic states $a_i$, $b_j$ are trapped in a shallow/deep potential and connected through a two-photon Raman transition.}
	\label{fig4}
\end{figure}

To conclude this section, we show that a closer inspection of the DMRG results provides further insight into the structure of the mEBS. We interpret the impurity as a lattice site described by hard-core boson operator $a_{\rm hc}$ and compute the two-point correlation functions
\begin{equation}
\label{eq:second}
G_{jm} = \bra{B_N} a^{\dagger}_{j} a_{m} \ket{B_N}
\end{equation}
where $j,m$ run over all the bath sites as well as the impurity site. It is found that the eigenvalues of the matrix $G_{jm}$ have only two dominant eigenvalues $p_{\pm}$ in all parameters regime and for all values of $N$ (see Fig.~\ref{fig4} for some examples). This implies that the mEBS $\ket{B_N}$ mainly lives in the symmetric space defined by two orthogonal modes. Without loss of generality, these two modes can be chosen as $A_1=A$ and $A_2=a_{\rm hc}\cos\theta+B\sin\theta$ and the bound state can then be approximated as
\begin{equation}
\ket{B_N} = \sum^{N}_{n_1=0} \sum^{N-n_1}_{n_2=0} \alpha_{n_1,n_2} \frac{(A^{\dagger}_{1})^{n_1}(A^{\dagger}_{2})^{n_2}}{\sqrt{n_1!n_2!}} \ket{0}
\end{equation}
The hard-core nature of the impurity imposes the constraint that $\alpha_{n_1,n_2}=0$ for $n_2>1$, which recovers the variational ansatz given in Eqs.~(\ref{eqmain:ansatz}). Indeed, the two eigenvalues computed from Eqs.~(\ref{eqmain:ansatz}) are $p_{\pm}=[N\pm\sqrt{N^{2}-4(N-1)(1-\beta ^{2})^{2}}]/2$.

\section{Preparation and detection of multiparticle bound states\label{sec:implementations}}

We have considered a very general model of a single impurity coupled to a bosonic bath which may be implemented in a plethora of different systems, ranging from superconducting circuits \cite{mlynek14a}, atoms near photonic crystals \cite{john90a,douglas15a,gonzaleztudela15b,goban13a,yu14a,lodahl13a,laucht12a,arcari14a,thompson13a} or cold atoms in optical lattices \cite{devega08a,navarretebenlloch11a}. In this Section, we show how to prepare and detect them in two of these platforms:

\emph{Optical emiters coupled to photonic reservoirs.} 
The observation of bound states in optical plaforms is challenging because typically the couplings are much smaller than the associated bandwidth ($\Omega\ll W$) and losses of the photons ($\kappa$) or the excited state of the impurity ($\Gamma$) which give rise to a finite lifetime of the mEBS. One option will be to use multiphoton scattering states, which was explored with single photon in Refs.~\cite{zhou08a,shi09a,shi09b,kocabacs15a,schneider15a,sanchezburillo14a}. Another way of circumventing these limitations is using atomic detection and post-selection at the expense of making the process probabilistic. The procedure to prepare a given $\ket{B_N}$ is to apply a sequence of $\pi$-pulses on the impurity and perform post-selection by projecting in the atomic state $P_s=\ket{s}\bra{s}$ with $s=g,e$. We start with the impurity in the ground state and no photons, i.e., $\ket{\Psi_0}=\ket{g}\otimes\ket{0_k}$ and focus on a situation with $\Delta<0$ and $|\Delta|\gg 1$. Then, if we apply a $\pi$-
pulse on the impurity and let it evolve, the atom will mainly 
follow $\ket{\Psi_0}\rightarrow \ket{e}\otimes \ket{0_k}\rightarrow \ket{B_1}$, where the contribution of scattering states will be very small because we are in a region where there are no modes to decay into. Then, we measure the impurity state $P_g$,  and apply another $\pi$-pulse if we detect an excitation, this will induce the change to $P_g\ket{B_1}\rightarrow \ket{e}\otimes \ket{1_k}$. As the state $\ket{e}\otimes \ket{1_k}$ is within the two-excitation subspace, it evolves to $\ket{B_2}$. To continue building up $\ket{B_N}$, we need to apply the sequence of $P_g$ measurements and $\pi$-pulse $N-1$ times to arrive at the desired excitation number $N$. Obviously, these sequences of $\pi$-pulses and post-selections must be faster than the lifetime determined by $\Gamma,\kappa$ such that the mEBS survives at the end of the process. For this reason, circuit QED \cite{mlynek14a}, though in the microwave regime, can be a better platform to observe the mEBS because both photon and qubit lifetimes are longer than those at optical setups.

\emph{Cold atoms in state dependent optical lattices.} Due to the limitations of standard quantum optical setups, one can think of using cold-atoms trapped in optical lattices to simulate this kind of Hamiltonians as originally proposed in Refs.~\cite{devega08a,navarretebenlloch11a}. This can be done by using state-dependent optical lattices in which two atomic states experience very different trapping potentials that can be obtained, e.g., using Alkaline-Earth atoms like Ytterbium~\cite{scazza14a,hofer15a}. Designing these lattices in a way such that the two atomic states experiences a very shallow/deep potential respectively. The ones in the shallow potential ($a_j$) have large tunneling amplitude $J$ and play the role of the bosonic bath; on the contrary, the ones in the deep potential will be localized and serve as the impurity ($b_j$). Moreover, we can tune the on-site interaction $U$ of the impurity states to $\infty$ such that it behaves effectively as a two-level system. The coupling between the two may be achieved through an off-resonant Raman transition to a common excited state yielding an effective 
number-conserving Hamiltonian. Interestingly, the coupling $\Omega$ and detuning $\Delta$ can be controlled independently of $J$ by the Raman parameters which allows us to explore the whole parameter space in Fig.~\ref{fig1}(b). Moreover, using the recently developed single-atom resolution microscopy and addressing techniques, we can achieve the single impurity regime and thus we can simulate the Hamiltonian $H$ \cite{devega08a,navarretebenlloch11a}. One way of preparing the $\ket{B_N}$ will be to start from a situation where the tunneling $J$ are switched off by increasing potential depth of the bath atoms and setting $\Delta$ to zero. Then, we load $N$ atoms in the bath mode coupled to the impurity, i.e., $a_0$, to give an initial state $\ket{\Psi_0}=\frac{(a_0^\dagger)^N}{\sqrt{N!}}\ket{0}\ket{g}=\ket{N}\ket{g}$. The protocol to prepare $\ket{B_N}$ consists of switching a strong Raman field with a $\pi$-phase for a time $t_N=\frac{3\pi}{\sqrt{n}\Omega}$ to reach the state $\ket{\Psi(t_N)}=\frac{1}{\sqrt{2}}(\ket{N-1}\ket{e}-\ket{N}\ket{g})$. After that,  we suddenly change the phase of $\Omega$ to be real such that $\ket{\Psi(t_N)}$ which coincides with $\ket{B_N}$ in the limit of $\Omega\gg |\Delta|,J$. Once we have such a state, we can change the state adiabatically with $\Delta$ or $J$ to explore the whole parameter space.

\section{Conclusions\label{sec:conclusions}}

To sum up, we unravel the existence of multiple excitation bound states in a very fundamental model of a single impurity coupled to a bosonic bath through a number conserving interaction. We first show that in certain regimes the impurity-bath coupling gives rise to an effective potential that is able to localize the particles around the impurity. Moreover, we provide the theoretical tools to characterize the bound states in all the parameter regimes by introducing a variational wavefunction which works for all energy dispersions and spatial dimensions. We test our variational ansatz for the case of a one-dimensional tight-binding model with exact and numerical calculations up to $N=5$ excitations and are able to distinguish different regimes depending on the scaling behavior of the energies or localization lengths, including one with non-analytical relations. Finally, we present two state-of-the-art implementations where the preparation and detection of these bound states is promising.

\section*{Acknowledgements}

The work was funded by the European Union Integrated project \emph{Simulators and Interfaces with Quantum Systems} (SIQS). AGT also acknowledges support from Alexander Von-Humboldt Foundation and Intra-European Marie-Curie Fellowship NanoQuIS (625955). During the preparation of the manuscript we became aware of a related recent preprint by Calajo \emph{et al} \cite{calajo15a} which also discusses the emergence of multiphoton bound states in waveguide QED setups. 

\vspace{-4mm}

\newpage \widetext

\begin{center}
\textbf{\large Appendix}
\end{center}

\setcounter{equation}{0} \setcounter{figure}{0} \setcounter{section}{0} 
\makeatletter

\renewcommand{\thefigure}{A\arabic{figure}} 
\renewcommand{\thesection}{A\arabic{section}} 
\renewcommand{\theequation}{A\arabic{equation}}

The supplemental material contains full details and calculations related to: i) Conditions for the existence of mEBS in Section~\ref{secSM:conditions}; ii) exact analytical solutions for $N=2$ and $N=3$ in Section \ref{secSM:exact}; iii) the optimization process of the variational ansatz in Section \ref{secSM:variational};

\section{Conditions for the existence of multiparticle bound states \label{secSM:conditions}}

The effective Hamiltonian in Eq.~(\ref{eq:effham}) can be obtained by applying the Fr\"{o}hlich transformation with generating function $S$
\begin{equation}
S = \frac{\Omega}{\sqrt{V}} \sum_{\mathbf{k}} \left( \frac{\eta_{\kk}a^{\dagger}_{\kk}\sigma_{ge}}{\Delta-\varepsilon_{\kk}} - \frac{\eta^{\ast}_{\kk}a_{\kk}\sigma_{eg}}{\Delta-\varepsilon_{\kk}} \right)
\end{equation}
on the original Hamiltonian. The transformed Hamiltonian $H_{\rm eff}=e^{-S}He^{S}$ is computed to the first order of $\Omega$ and results in Eq.~(\ref{eq:effham}). The effective Hamiltonian can be projected to the $\ket{e}$ ($\ket{g}$) state in the PG (PE) regime to give $H^{s}_{\rm eff}$ ($s=e$ or $g$) as
\begin{equation}
H^{s}_{\rm eff} = E_{1} + H_{\rm bath} - \frac{\Omega^{2}}{2V} \sum_{\kk^{\prime}} \eta^{\ast}_{\kk^{\prime}} \eta_{\kk} \left( \frac{1}{|\Delta-\varepsilon_{\kk^{\prime}}|} + \frac{1}{|\Delta-\varepsilon_{\kk}|} \right) a^{\dagger}_{\kk} a_{\kk^{\prime}}  
\label{HI}
\end{equation}
The projected effective Hamiltonian defines a secular equation $H^{s}_{\rm eff}A^{\dagger}_{s}\ket{0}=E^{s}_{1}A^{\dagger}_{s}\ket{0}$. The operator $A_{s}$ is a collective bath operator which can be expanded as $\sum_{\kk} \varphi^{s}_{A}(\kk) a_{\kk}$. The bath component $\ket{\Psi_{e}}=\frac{A^{\dagger{N-1}}_{e}\ket{0}}{\sqrt{(N-1)!}}$ in the PG regime, whereas the bath component $\ket{\Psi_{g}}=\frac{A^{\dagger{N}}_{g}\ket{0}}{\sqrt{N!}}$ in the PE regime. 

The ground state energy of the mEBS is $E^{g}_{N}=E_{1}+(N-1)E^{g}_{1}$ ($E_1$ is the energy of sEBS defined in the main text) in the PG regime and $E^{e}_{N}=NE^{e}_{1}$ in the PE regime. The coefficients in the expansion of $A_{s}$ are
\begin{equation}
\varphi^{s}_{A} (\kk) = \frac{\Omega^{2}\eta_{\kk}}{2(E^{s}_{1}-\varepsilon_{\kk})} \left( C_{2} - \frac{C_{1}}{|\Delta-\varepsilon_{\kk}|} \right)
\end{equation}
The variables $C^s_{1}$ and $C^s_{2}$ can be obtained by solving the self-consistent equation $\mathbf{M}^s (E^{s}_{1},\Delta) \mathbf{C}^s=0$. $\mathbf{M}^s$ is a $2{\times}2$ matrix with elements
\begin{eqnarray}
&& \mathbf{M}^{s}_{11}(E,\Delta)=\mathbf{M}^{s}_{22}(E,\Delta)=1+\frac{w_1(E)-w_1(\Delta)}{|\Delta-E|} \nonumber \\
&& \mathbf{M}^{s}_{12}(E,\Delta)=-w_1(E) \;\;\; \mathbf{M}^{s}_{21}(E,\Delta)=-\frac{w_1(E)-w_1(\Delta)}{(\Delta -E)^{2}}-\frac{\partial_{\Delta}w_1(\Delta )}{\Delta-E}
\end{eqnarray}
where
\begin{equation}
w_\alpha(E) = \frac{\Omega^{2}}{2} \int\frac{d^d\kk}{(2\pi)^{d}} \frac{\varepsilon_\kk^{\alpha-1} |\eta_{\kk}|^{2}}{(E-\varepsilon_{\kk})^\alpha}
\end{equation}
and $\mathbf{C}=(C_{1},C_{2})^{T}$. 

The mEBS exist if the equation $F^s_N(E)={\rm det}\mathbf{M^{s}}(E^{s})=0$ has solutions $E^{s}_1<0$, {\em i.e.}, they lie outside of the bath spectrum. Because $F^s_N(E)$ is a continuous function and is negative when $E\rightarrow\infty$, there will be at least one solution if $F^{s}_{N}(0)<0$. From the explicit expression of $F^s_N$
\begin{eqnarray}
F^{s}_{N}(0) = \left[1-\frac{w_1(\Delta)}{|\Delta|}\right]^{2} -\left[2+\frac{w_2(\Delta)}{|\Delta|}\right]\frac{\Omega^{2}I_{0}}{2|\Delta|} < 0
\end{eqnarray}
we can see that the existence or absence of mEBS depends on $I_{0}$. If $I_{0}\rightarrow+\infty$ due to infrared divergence, we have $F^{s}_{N}(0)\rightarrow -\infty$ and mEBS exist in both regimes. The opertor $C^{\dagger}_e$ and the number $D_g$ that we mentioned in the main text are 
\begin{eqnarray}
&& C^{\dagger}_e = \frac{\Omega}{\sqrt{V}} \sum_{\kk} \frac{\eta_{\kk}a^{\dagger}_{\kk}}{\Delta-\varepsilon_{\kk}} \\
&& D_g = \frac{\Omega\sqrt{N}}{\sqrt{V}} \sum_{\kk} \frac{\eta^{\ast}_{\kk}\varphi_{A}(\kk)}{\Delta-\varepsilon_{\kk}}
\end{eqnarray}

\section{Exact results for $N=2$ and $N=3$ \label{secSM:exact}}

In this Section, we provide the exact results for the mEBS in the $N=2$ and $3$ subspaces using quantum field theory. For convenience, we introduce a hard-core boson described by the annihilation (creation) operator $a_{\rm hc}{\equiv}b_{0}$ ($a^{\dagger}_{\rm hc}{\equiv}b^{\dagger}_{0}$) and rewrite the Hamiltonian as $H=H_{0}+H_{\rm hc}$ where $H_{0}$ is a quadratic term
\begin{equation}
H_{0} = H_{\rm bath} + \Delta b^{\dagger}_{0} b_{0} + \frac{\Omega}{\sqrt{V}} \sum_{\kk} \eta_{\kk} (a^{\dagger}_{\kk} b_{0} + b^{\dagger}_{0} a_{\kk} )
\end{equation}
and $H_{\rm hc}$ represents a hard-core interaction
\begin{equation}
H_{\rm hc} = \frac{U_{0}}{2} b^{\dagger2}_{0} b^{2}_{0}
\end{equation}
with an infinite strength $U_{0}$. The total particle number $N=b_{0}^{\dagger
}b_{0}+\sum_{\mathbf{k}}a_{\mathbf{k}}^{\dagger }a_{\mathbf{k}}$ is still a conserved quantity, so we can diagonalize the Hamiltonian separately in the subspaces with different $N$. For example, the eigenstate for $N=1$ has the form
\begin{equation}
\left\vert 1_{\lambda }\right\rangle =[u_{\lambda }b_{0}^{\dagger }+\sum_{\mathbf{k}}f_{\lambda }(\mathbf{k})a_{\mathbf{k}}^{\dagger }]\left\vert
0\right\rangle ,
\end{equation}
where $E_{1\lambda}$ is the energy eigenvalue. The secular equation is a set of coupled equations
\begin{eqnarray}
\Delta u_{\lambda }+\frac{\Omega }{\sqrt{V}}\sum_{\mathbf{k}}\eta
_{\mathbf{k}}f_{\lambda }(\mathbf{k}) &=&E_{1\lambda }u_{\lambda }  \notag \\
\varepsilon _{\mathbf{k}}f_{\lambda }(\mathbf{k})+\frac{\Omega }{\sqrt{
V}}\eta _{\mathbf{k}}u_{\lambda } &=&E_{1\lambda }f_{\lambda }(\mathbf{k}) 
\label{sb}
\end{eqnarray}

For the bound state energy $E_{1}\notin \varepsilon _{\mathbf{k}}$ outside the continuum $\varepsilon_{\mathbf{k}}$, solving Eq.~(\ref{sb}) gives the wavefunction
\begin{equation}
f_{\mathrm{B}}(\mathbf{k})=\frac{\Omega }{\sqrt{V}}\frac{\eta _{\mathbf{k}}u_{\mathrm{B}}}{E_{1}-\varepsilon _{\mathbf{k}}}
\end{equation}
where the normalization factor is
\begin{equation}
u_{\mathrm{B}}^{-2}=1+\int \frac{d^{d}\kk}{(2\pi )^{d}}\frac{\Omega ^{2}\eta _{\mathbf{k}}^{2}}{(E_{1}-\varepsilon _{k})^{2}},
\end{equation}
and the bound state energy satisfies the equation
\begin{equation}
E_{1}=\Delta +\int \frac{d^{d}\kk}{(2\pi )^{d}}\frac{\Omega ^{2}\eta_{\mathbf{k}}^{2}}{E_{1}-\varepsilon_{\mathbf{k}}}.
\label{E1b}
\end{equation}

For a general dispersion relation $\varepsilon_{\mathbf{k}}$, the spectrum $E_{1\lambda} $ is obtained by numerical diagonalization of
\begin{equation}
\mathcal{H}_{0}=\left(
\begin{array}{cccc}
\Delta & \frac{\Omega }{\sqrt{V}}\eta _{\mathbf{k}_{1}} & ... &
\frac{\Omega }{\sqrt{V}}\eta _{\mathbf{k}_{N}} \\
\frac{\Omega }{\sqrt{V}}\eta _{\mathbf{k}_{1}} & \varepsilon_{\mathbf{k}_{1}} & 0 & 0 \\
... & 0 & ... & 0 \\
\frac{\Omega }{\sqrt{V}}\eta _{\mathbf{k}_{N}} & 0 & 0 &
\varepsilon _{\mathbf{k}_{N}}
\end{array}
\right),
\label{Hs0}
\end{equation}
whose eigenvectors determine the parameters $u_{\lambda}$ and the wavefunction $f_{\lambda}(\mathbf{k})$. 

To analyze the mEBS, it is convenient to introduce the Green functions
\begin{eqnarray}
G_{bb}^{(0)}(t) &\equiv &G_{b}^{(0)}(t)=-i\left\langle 0\right\vert \mathcal{T}b_{0}(t)b_{0}^{\dagger }(0)\left\vert 0\right\rangle, \notag \\
G_{a_{\mathbf{k}}b}^{(0)}(t) &=&-i\left\langle 0\right\vert \mathcal{T} a_{\mathbf{k}}(t)b_{0}^{\dagger }(0)\left\vert 0\right\rangle,
\end{eqnarray}
where $\mathcal{T}$ is the time-ordering operator. The Fourier transforms $G_{\alpha b}^{(0)}(\omega )=\int dtG_{\alpha b}^{(0)}(t)e^{i\omega t}$ ($\alpha =a_{\mathbf{k}},b_{0}$) can be obtained by integrating out the bath
modes as
\begin{eqnarray}
G_{a_{\mathbf{k}}b}^{(0)}(\omega ) &=&\frac{\Omega \eta _{\mathbf{k}}}{\sqrt{V}}\frac{G_{b}^{(0)}(\omega )}{\omega -\varepsilon _{\mathbf{k}}+i0^{+}}, \\
G_{b}^{(0)}(\omega ) &=&[\omega -\Delta -\int \frac{d^{d}\kk}{(2\pi )^{d}} \frac{\Omega ^{2}\eta _{\mathbf{k}}^{2}}{\omega -\varepsilon _{\mathbf{k}}+i0^{+}}]^{-1},
\end{eqnarray}
The poles of $G_{b}^{(0)}(\omega)$ correspond the bound state energies as shown in Eq.~(\ref{E1b}). 

In terms of the eigenstates and eigenvalues of $\mathcal{H}_{0}$, the Green function
\begin{equation}
G_{b}^{(0)}(\omega )=\sum_{\lambda }\frac{\left\vert u_{\lambda }\right\vert
^{2}}{\omega -E_{1\lambda }+i0^{+}}
\end{equation}
can be constructed using the Lehmann representation.

\subsection{The $N=2$ subspace}

In the $N=2$ subspace, the most general eigenstate with eigenenergy $E_{2\lambda}$ has the form
\begin{equation}
\left\vert 2_{\lambda }\right\rangle =[u_{\lambda }^{(2)}b_{0}^{\dagger2}+\sum_{\mathbf{k}}f_{1\lambda }^{(2)}(\mathbf{k})a_{\mathbf{k}}^{\dagger}b_{0}^{\dagger }+\sum_{\mathbf{k}_{1},\mathbf{k}_{2}}f_{2\lambda }^{(2)}(
\mathbf{k}_{1},\mathbf{k}_{2})a_{\mathbf{k}_{1}}^{\dagger }a_{\mathbf{k}_{2}}^{\dagger}]\left\vert 0\right\rangle\,.
\end{equation}

The goal is this Section is to obtain the bound state energy $E_{2}$, the coefficient $u^{(2)}_{\rm B}$, the bath wavefunctions $f_{1{\rm B}}^{(2)}(\mathbf{k})$ and $f_{2{\rm B}}^{(2)}(\mathbf{k}_{1},\mathbf{k}_{2})$ using the Green function method. The two-particle Green functions are defined as
\begin{equation}
G_{\alpha _{1}\alpha _{2}}(t)=-i\left\langle 0\right\vert \mathcal{T}\alpha_{1}(t)\alpha _{2}(t)\alpha _{2}^{\dagger }\alpha _{1}^{\dagger }\left\vert0\right\rangle ,
\end{equation}
where there are three different possible choices, i) $\alpha _{1}=\alpha _{2}=b_{0}$; ii) $\alpha
_{1}=b_{0}$, $\alpha _{2}=a_{\mathbf{k}}$; iii) $\alpha _{1}=a_{\mathbf{k}_{2}}$, $\alpha _{2}=a_{\mathbf{k}_{1}}$. In the Lehmann representation
\begin{equation}
G_{\alpha _{1}\alpha _{2}}(\omega )=\sum_{\lambda }\frac{\left\langle0\right\vert \alpha _{1}\alpha _{2}\left\vert 2_{\lambda }\right\rangle\left\langle 2_{\lambda }\right\vert \alpha _{2}^{\dagger }\alpha_{1}^{\dagger}\left\vert 0\right\rangle }{\omega -E_{2\lambda }+i0^{+}},
\end{equation}
the poles and the corresponding residues of the Fourier transform $G_{\alpha_{1}\alpha _{2}}(\omega )=\int dtG_{\alpha _{1}\alpha _{2}}(t)e^{i\omega t}$ determine the bound state energy $E_{2}$ and wavefunctions $\left\langle 0\right\vert \alpha _{1}\alpha _{2}\ket{B_2} $ ($u_{\mathrm{B}}^{(2)}=\left\langle 0\right\vert
b_{0}^{2}\ket{B_2} /2$, $f_{1\mathrm{B}}^{(2)}(\mathbf{k})=\left\langle 0\right\vert b_{0}a_{\mathbf{k}}\ket{B_2} $ and $f_{2\mathrm{B}}^{(2)}(\mathbf{k}_{1},\mathbf{k}_{2})=\left\langle 0\right\vert a_{\mathbf{k}_{2}}a_{\mathbf{k}_{1}}\ket{B_2} /2$).

The Dyson expansion of the two-body interaction $H_{\mathrm{hc}}$ results in the connected part $[G_{\alpha _{1}\alpha _{2}}(\omega )]_{\mathrm{c}}$ of two-body Green function
\begin{equation}
\lbrack G_{\alpha _{1}\alpha _{2}}(\omega )]_{\mathrm{c}}=2\Pi _{\alpha
_{1}\alpha _{2}}(\omega )T_{2}(\omega )\Pi _{\alpha _{1}\alpha _{2}}^{\ast
}(\omega ).
\end{equation}
The poles of the $T$-matrix
\begin{equation}
T_{2}(\omega )=\frac{1}{U_{0}^{-1}-\Pi _{bb}(\omega )}
\end{equation}
give the bound state energy, where
\begin{equation}
\Pi_{bb}(\omega )=\sum_{\lambda \lambda ^{\prime }}\frac{\left\vert{u_{\lambda}}\right\vert^{2}\left\vert u_{\lambda ^{\prime }}\right\vert ^{2}}{\omega -E_{1\lambda }-E_{1\lambda ^{\prime }}+i0^{+}}.
\end{equation}
can be computed from the bubble diagram. In the vicinity of a pole $E_2$, the $T$-matrix $T_{2}(\omega )\sim Z_{2\mathrm{B}}/(\omega -E_2+i0^{+})$ with a residue
\begin{equation}
Z_{2\mathrm{B}}^{-1}=\sum_{\lambda \lambda ^{\prime }}\frac{\left\vert u_{\lambda} \right\vert ^{2}\left\vert u_{\lambda ^{\prime}} \right\vert ^{2}}{(E_2-E_{1\lambda }-E_{1\lambda ^{\prime}})^{2}}.
\end{equation}

The wavefunctions can be determind by $\Pi_{\alpha _{1}\alpha_{2}}(\omega)$ which are
\begin{equation}
\Pi _{ba_{\mathbf{k}}}(\omega )=\frac{\Omega \eta _{\mathbf{k}}}{\sqrt{V}}\sum_{\lambda }\frac{\left\vert u_{\lambda }\right\vert^{2}G_{b}^{(0)}(\omega -E_{1\lambda })}{\omega -\varepsilon _{\mathbf{k}}-E_{1\lambda }+i0^{+}},
\end{equation}
and
\begin{eqnarray}
\Pi _{a_{\mathbf{k}_{2}}a_{\mathbf{k}_{1}}}(\omega ) &=& \frac{\Omega ^{2}}{V}\eta _{\mathbf{k}_{1}}\eta _{\mathbf{k}_{2}}\sum_{\lambda\lambda ^{\prime }}\frac{\left\vert u_{\lambda }\right\vert ^{2}\left\vert u_{\lambda ^{\prime }}\right\vert ^{2}}{(\omega -\varepsilon _{\mathbf{k}_{1}}-\varepsilon _{\mathbf{k}_{2}}+i0^{+})(\omega -E_{1\lambda}-E_{1\lambda ^{\prime }}+i0^{+})}  \notag \\
&& \times \frac{2\omega -\varepsilon _{\mathbf{k}_{1}}-\varepsilon _{\mathbf{k}_{2}}-E_{1\lambda }-E_{1\lambda ^{\prime }}}{(\omega -\varepsilon _{\mathbf{k}_{1}}-E_{1\lambda }+i0^{+})(\omega -E_{1\lambda ^{\prime }}-\varepsilon_{\mathbf{k}_{2}}+i0^{+})}
\end{eqnarray}

The variables appearing in the wavefunctions are $u_{\mathrm{B}}^{(2)}=\sqrt{Z_{2\mathrm{B}}/2}\Pi_{bb}(E_2)$, $f_{1\mathrm{B}}^{(2)}(\mathbf{k})=\sqrt{2Z_{2\mathrm{B}}}\Pi _{ba_{\mathbf{k}}}(E_2)$, and $f_{2\mathrm{B}
}^{(2)}(\mathbf{k}_{1},\mathbf{k}_{2})=\sqrt{Z_{2B}/2}\Pi _{a_{\mathbf{k}_{2}}a_{\mathbf{k}_{1}}}(E_2)$. The real space wavefunctions obtained via Fourier transforms are
\begin{eqnarray}
f^{(2)}_{1{\rm B}}(\mathbf{r}_j) = \frac{1}{\sqrt{V}} \sum_{\mathbf k} f_{1{\rm B}}^{(2)}(\mathbf{k}) e^{i\mathbf{k}\cdot\mathbf{r}_j} \;\;\; f^{(2)}_{2{\rm B}}(\mathbf{r}_{j_1},\mathbf{r}_{j_2}) = \frac{1}{V} \sum_{\mathbf{k}_{1}\mathbf{k}_{2}} f_{2\mathrm{B}}^{(2)}(\mathbf{k}_{1},\mathbf{k}_{2}) e^{i\mathbf{k}_{1}\cdot\mathbf{r}_{j_1}+i\mathbf{k}_{2}\cdot\mathbf{r}_{j_2}}
\end{eqnarray}
In the hard-core limit $U_{0}\rightarrow \infty $, $T_{2}^{-1}(E_2)=0$ gives $\Pi_{b}(E_2)=0$ so $u_{\mathrm{B}}^{(2)}=0$, which means that there is no double occupation in the $b_0$ mode.

\subsection{The $N=3$ subspace}

In the $N=3$ subspace, the most general eigenstate reads
\begin{eqnarray}
\left\vert 3_{\lambda }\right\rangle &=&[u_{\lambda }^{(3)}b_{0}^{\dagger3}+\sum_{\mathbf{k}}f_{1\lambda }^{(3)}(\mathbf{k})a_{\mathbf{k}}^{\dagger}b_{0}^{\dagger 2}+\sum_{\mathbf{k}_{1}\mathbf{k}_{2}}f_{2\lambda }^{(3)}(\mathbf{k}_{1},\mathbf{k}_{2})a_{\mathbf{k}_{1}}^{\dagger }a_{\mathbf{k}_{2}}^{\dagger }b_{0}^{\dagger } + \sum_{\mathbf{k}_{1}\mathbf{k}_{2}\mathbf{k}_{3}}f_{3\lambda }^{(3)}(\mathbf{k}_{1},\mathbf{k}_{2},\mathbf{k}_{3})a_{\mathbf{k}_{1}}^{\dagger }a_{\mathbf{k}_{2}}^{\dagger }a_{\mathbf{k}_{3}}^{\dagger }]\left\vert0\right\rangle
\end{eqnarray}
The three-particle bound state can be studied by the three-particle Green function
\begin{equation}
G_{3}(t)=-i\left\langle \mathcal{T}\alpha _{3}(t)\alpha _{2}(t)\alpha_{1}(t)\alpha _{1}^{\dagger }\alpha _{2}^{\dagger }\alpha _{3}^{\dagger}\right\rangle ,
\end{equation}
where the bound state energy $E_3$, the coefficient $u_{\mathrm{B}}^{(3)}$, the wavefunctions $f_{1\mathrm{B}}^{(3)}(\mathbf{k})$, $f_{2\mathrm{B}}^{(3)}(\mathbf{k}_{1},\mathbf{k}_{2})$, and $f_{3\mathrm{B}}^{(3)}(\mathbf{k}_{1},\mathbf{k}_{2},\mathbf{k}_{3})$ can be obtained by four different choices of operators in $G_{3}(t)$: i) $\alpha _{1}=\alpha _{2}=\alpha _{3}=b_{0}$; ii) $\alpha _{1}=a_{\mathbf{k}}$, $\alpha _{2}=\alpha _{3}=b_{0}$; iii) $\alpha _{1}=a_{\mathbf{k}_{1}}$, $\alpha _{2}=a_{\mathbf{k}_{2}}$, $\alpha_{3}=b_{0}$, and iv) $\alpha _{1}=a_{\mathbf{k}_{1}}$, $\alpha _{2}=a_{\mathbf{k}_{2}}$, $\alpha _{3}=a_{\mathbf{k}_{3}}$. 

In the Lehmann representation
\begin{equation}
G_{3}(\omega) = \int dtG_{3}(t)e^{i\omega t} = \sum_{\lambda }\frac{\left\langle 0\right\vert \alpha_{3}\alpha _{2}\alpha _{1}\left\vert 3_{\lambda }\right\rangle \left\langle 3_{\lambda }\right\vert \alpha _{1}^{\dagger }\alpha _{2}^{\dagger }\alpha
_{3}^{\dagger }\left\vert 0\right\rangle }{\omega -E_{3\lambda }+i0^{+}},
\end{equation}
the poles and residues determine the bound state energy $E_3$ and the wavefunctions $\left\langle 0\right\vert \alpha_{3}\alpha _{2}\alpha _{1}\ket{B_3} $ ($u_{\mathrm{B}}^{(3)}=\left\langle 0\right\vert b_{0}^{3}\left\vert 3_{\lambda}\right\rangle /6$, $f_{1\mathrm{B}}^{(3)}(\mathbf{k})=\left\langle0\right\vert b_{0}^{2}a_{\mathbf{k}}\ket{B_3} /2$, $f_{2\mathrm{B}}^{(3)}(\mathbf{k}_{1},\mathbf{k}_{2})=\left\langle 0\right\vert b_{0}a_{\mathbf{k}_{2}}a_{\mathbf{k}_{1}}\ket{B_3} /2$, and $f_{3\lambda }^{(3)}(\mathbf{k}_{1},\mathbf{k}_{2},\mathbf{k}_{3})=\left\langle 0\right\vert a_{\mathbf{k}_{3}}a_{\mathbf{k}_{2}}a_{\mathbf{k}_{1}}\ket{B_3}/6$).

The Dyson expansion gives to the connected part
\begin{equation}
\lbrack G_{3}(\omega )]_{\mathrm{c}}=\int \frac{d\omega _{1}d\omega
_{1}^{\prime }}{(2\pi )^{2}}\chi _{3}(\omega ,\omega _{1})T_{3}(\omega
,\omega _{1},\omega _{1}^{\prime })\chi _{3}(\omega ,\omega _{1}^{\prime}),
\end{equation}
where
\begin{eqnarray}
\chi _{3}(\omega ,\omega _{1}) &=& P_{123}[G_{\alpha _{i_{1}}b}^{(0)}(\omega
_{1})T_{2}(\omega -\omega _{1}) \int \frac{d\omega _{2}}{2\pi }G_{\alpha _{i_{2}}b}^{(0)}(\omega
_{2})G_{\alpha _{i_{3}}b}^{(0)}(\omega -\omega _{1}-\omega _{2})],
\end{eqnarray}
is determined by the operator $P_{123}$ which permutes $\alpha _{1,2,3}$. The three particle $T$-matrix satisfies the integral equation
\begin{eqnarray}
T_{3}(\omega ,\omega _{1},\omega _{1}^{\prime }) &=& G_{b}^{(0)}(\omega-\omega_{1}-\omega _{1}^{\prime })+2i\int \frac{d\omega^{\prime}}{2\pi} G_{b}^{(0)}(\omega -\omega _{1}-\omega ^{\prime }) T_{2}(\omega -\omega ^{\prime })G_{b}^{(0)}(\omega ^{\prime})T_{3}(\omega ,\omega ^{\prime },\omega _{1}^{\prime }).
\end{eqnarray}

In the vicinity of the pole $E_3$, the three-particle $T$-matrix has the form
\begin{equation}
T_{3}(\omega ,\omega _{1},\omega _{1}^{\prime })=\frac{F(\omega
_{1})F(\omega _{1}^{\prime })}{\omega -E_3+i0^{+}},
\end{equation}
where $F(\omega )$ is determined by the integral equation
\begin{equation}
F(\omega )=2i\int \frac{d\omega ^{\prime }}{2\pi }G_{b}^{(0)}(E_{3}-\omega -\omega ^{\prime })T_{2}(E_3-\omega ^{\prime})G_{b}^{(0)}(\omega ^{\prime })F(\omega ^{\prime }).
\label{Int}
\end{equation}
Using the analyticity of $F(\omega)$ and the residue theorem, we find that the integral in Eq.~(\ref{Int})
becomes
\begin{equation}
F(\omega)=2\sum_{\lambda }\left\vert u_{\lambda}\right\vert^{2}G_{b}^{(0)}(E_3-\omega-E_{1\lambda})T_{2}(E_{3}-E_{1\lambda})F(E_{1\lambda }).
\label{Fs}
\end{equation}
For on-shell frequency $\omega=E_{1\lambda }$, we obtain the matrix equation
\begin{equation}
\sum_{\lambda^{\prime}}M_{\lambda\lambda^{\prime}}F(E_{1\lambda^{\prime}})=0
\end{equation}
where
\begin{equation}
M_{\lambda\lambda^{\prime}}=2|u_{\lambda ^{\prime }}|^{2} G_{b}^{(0)}(E_3-E_{1\lambda }-E_{1\lambda ^{\prime}}) T_{2}(E_3-E_{1\lambda^{\prime }})-\delta_{\lambda\lambda^{\prime}}.
\end{equation}
The bound state energy $E_3$ is obtained by solving the equation $\det M=0$. The eigenvector $F(E_{1\lambda })$ with zero eigenvalue can be used to obtain the function $F(\omega )$ via Eq.~(\ref{Int}).

In the hard-core limit $U_{0}\rightarrow\infty$, the wavefunctions $u^{(3)}_{\rm B}=f^{(3)}_{1{\rm B}}(\mathbf{k})=0$ vanish, and the residue of $G_{3}(\omega)$ gives the wavefunctions
\begin{eqnarray}
f_{2\mathrm{B}}^{(3)}(\mathbf{k}_{1},\mathbf{k}_{2}) &=& -\frac{\Omega^{2}}{V}\eta _{\mathbf{k}_{1}}\eta _{\mathbf{k}_{2}} P_{12} \Bigg\{ \sum_{\lambda_{1}\lambda _{2}\lambda _{3}}\frac{\left\vert u_{\lambda _{1}}\right\vert^{2}\left\vert u_{\lambda _{2}}\right\vert ^{2}\left\vert u_{\lambda_{3}}\right\vert^{2}}{2(\varepsilon _{\mathbf{k}_{1}}-E_{1\lambda _{1}})} \nonumber \\
&\times& \Bigg[ \frac{T_{2}(E_3-E_{1\lambda _{2}})F(E_{1\lambda_{2}})+2T_{2}(E_3-\varepsilon _{\mathbf{k}_{1}})F(\varepsilon _{\mathbf{k}_{1}})}{(E_3-\varepsilon _{\mathbf{k}_{1}}-\varepsilon_{\mathbf{k}_{2}}-E_{1\lambda _{2}})(E_3-\varepsilon _{\mathbf{k}_{1}}-E_{1\lambda _{2}}-E_{1\lambda _{3}})} \nonumber \\
&& -\frac{T_{2}(E_3-E_{1\lambda _{2}})F(E_{1\lambda_{2}})+2T_{2}(E_3-E_{1\lambda _{1}})F(E_{1\lambda _{1}})}{(E_3-\varepsilon _{\mathbf{k}_{2}}-E_{1\lambda _{1}}-E_{1\lambda_{2}})(E_3-E_{1\lambda _{1}}-E_{1\lambda _{2}}-E_{1\lambda _{3}})} \Bigg] \Bigg\}
\end{eqnarray}
and
\begin{eqnarray}
f_{3\mathrm{B}}^{(3)}(\mathbf{k}_{1},\mathbf{k}_{2},\mathbf{k}_{3}) &=& -\frac{\Omega^{3}}{6V^{3/2}} \eta_{\mathbf{k}_{1}}\eta _{\mathbf{k}_{2}}\eta_{\mathbf{k}_{3}} P_{123} \Bigg\{ \sum_{\lambda _{1}\lambda_{2}\lambda _{3}}\frac{\left\vert u_{\lambda _{1}}\right\vert ^{2}\left\vert u_{\lambda _{2}}\right\vert ^{2}\left\vert u_{\lambda _{3}}\right\vert ^{2}}{\varepsilon _{\mathbf{k}_{1}}-E_{1\lambda _{1}}} \nonumber \\
&\times& \Bigg[ \frac{T_{2}(E_3-\varepsilon _{\mathbf{k}_{1}})F(\varepsilon_{\mathbf{k}_{1}})}{(E_3-\varepsilon _{\mathbf{k}_{1}}-\varepsilon_{\mathbf{k}_{2}}-\varepsilon _{\mathbf{k}_{3}})(E_3-\varepsilon_{\mathbf{k}_{1}}-\varepsilon _{\mathbf{k}_{2}}-E_{1\lambda _{3}})} \nonumber \\
&\times& \frac{2E_3-2\varepsilon _{\mathbf{k}_{1}}-\varepsilon_{\mathbf{k}_{2}}-\varepsilon _{\mathbf{k}_{3}}-E_{1\lambda _{2}}-E_{1\lambda_{3}}}{(E_3-\varepsilon _{\mathbf{k}_{1}}-\varepsilon _{\mathbf{k}_{3}}-E_{1\lambda _{2}})(E_3-\varepsilon _{\mathbf{k}_{1}}-E_{1\lambda _{2}}-E_{1\lambda _{3}})} \nonumber \\
&-& \frac{T_{2}(E_3-E_{1\lambda _{1}})F(E_{1\lambda _{1}})}{(E_3-\varepsilon _{\mathbf{k}_{2}}-\varepsilon _{\mathbf{k}_{3}}-E_{1\lambda _{1}})(E_3-\varepsilon _{\mathbf{k}_{2}}-E_{1\lambda _{1}}-E_{1\lambda _{3}})} \nonumber \\
&\times& \frac{2E_3-\varepsilon _{\mathbf{k}_{2}}-\varepsilon _{\mathbf{k}_{3}}-2E_{1\lambda _{1}}-E_{1\lambda _{2}}-E_{1\lambda _{3}}}{(E_{3}-\varepsilon _{\mathbf{k}_{3}}-E_{1\lambda _{1}}-E_{1\lambda _{2}})(E_3-E_{1\lambda _{1}}-E_{1\lambda _{2}}-E_{1\lambda _{3}})} \Bigg] \Bigg\}
\end{eqnarray}
where $P_{12}$ and $P_{123}$ permutation for $\mathbf{k}_{1},\mathbf{k}_{2}$ and $\mathbf{k}_{1},\mathbf{k}_{2},\mathbf{k}_{3}$, respectively. The real space wavefunctions can be obtained via Fourier transforms
\begin{eqnarray}
f^{(3)}_{2{\rm B}}(\mathbf{r}_{j_1},\mathbf{r}_{j_2}) &=& \frac{1}{V} \sum_{\mathbf{k}_{1}\mathbf{k}_{2}} f^{(3)}_{2{\rm B}} (\mathbf{k}_{1},\mathbf{k}_{2}) e^{i\mathbf{k}_{1}\cdot\mathbf{r}_{j_1}+i\mathbf{k}_{2}\cdot 
\mathbf{r}_{j_2}}, \notag \\
f^{(3)}_{3{\rm B}}(\mathbf{r}_{j_1},\mathbf{r}_{j_2},\mathbf{r}_{j_3}) &=& \frac{1}{V^{3/2}} \sum_{\mathbf{k}_{1}\mathbf{k}_{2}\mathbf{k}_{3}} f^{(3)}_{3{\rm B}} (\mathbf{k}_{1},\mathbf{k}_{2},\mathbf{k}_{3}) e^{i\mathbf{k}_{1}\cdot\mathbf{r}_{j_1}+i\mathbf{k}_{2}\cdot\mathbf{r}_{j_2}+i\mathbf{k}_{3}\cdot\mathbf{r}_{j_3}}
\end{eqnarray}

\section{Variational Ansatz\label{secSM:variational}}

The ground state energy of the variational ansatz Eq.~(\ref{eqmain:ansatz}) is given by
\begin{equation}
E_{\mathrm{GS}}=(N-1)h_{AA}+v^{T}\mathcal{E}_{0}v,
\label{Egs}
\end{equation}
where we have used a matrix
\begin{equation}
\mathcal{E}_{0}=
\left(
\begin{array}{ccc}
\Delta & \sqrt{N}\Omega I_{A} & \Omega I_{B} \\
\sqrt{N}\Omega I_{A} & h_{AA} & \sqrt{N}h_{AB} \\
\Omega I_{B} & \sqrt{N}h_{AB} & h_{BB}
\end{array}
\right) 
\label{E0}
\end{equation}
with variables $h_{s_{1}s_{2}}=\sum_{\kk} \varepsilon_{\kk} \varphi_{s_{1}}(\kk) \varphi_{s_{2}}(\kk)$ and $I_{s}=\sum_{\kk}\eta_{\kk}\varphi_{s}(\kk)/\sqrt{V}$. The orthogonal collective modes used in the main text satisfy the conditions $[A,A^{\dagger}]=[B,B^{\dagger }]=1$ and $[A,B^{\dagger }]=0$, which gives
\begin{equation}
\sum_{\kk} \varphi^{2}_{s}(\kk)=1 \;\;\;{\rm and} \;\;\; \sum_{\kk} \varphi_{A}(\kk) \varphi_{B} (\kk) = 0
\label{C2}
\end{equation}
To optimize the ground state energy under these constraints, we introduce Lagrangian multipliers $\lambda$, $\mu_{A}$, $\mu_{B}$, and $\mu _{AB}$ to define the function 
\begin{eqnarray}
F_{\rm GS} &=& E_{\rm GS} - \mu _{A} \sum_{\kk} \varphi^{2}_{A}(\kk) - \mu_{B} \sum_{\kk} \varphi^{2}_{B}(\kk) -2\mu_{AB} \sum_{\kk} \varphi_{A}(\kk) \varphi_{B} (\kk) - \lambda (\alpha^{2}+\beta^{2}+\gamma^{2})
\end{eqnarray}

By taking the derivative to respect to $v$ and $\varphi _{s}(\mathbf{k})$, we obtain a set of coupled nonlinear Gross-Pitaevski-like equations
\begin{equation}
\mathcal{E}_{0}v=\lambda v,
\label{GP1}
\end{equation}
and
\begin{equation}
\mathcal{H}_{\rm GP}\left(
\begin{array}{c}
\varphi _{A}(\kk) \\
\varphi _{B}(\kk)
\end{array}
\right) + \frac{\Omega \eta_{\kk}}{\sqrt{V}} \alpha 
\left(
\begin{array}{c}
\sqrt{N}\beta \\
\gamma
\end{array}
\right) = \mu 
\left(
\begin{array}{c}
\varphi _{A}(\kk) \\
\varphi _{B}(\kk)
\end{array}
\right)
\label{GP2}
\end{equation}
where
\begin{equation}
\mathcal{H}_{\rm GP}=\left(
\begin{array}{cc}
(N-1+\beta^{2})\varepsilon_{\mathbf{k}} & \beta \gamma \sqrt{N} \varepsilon_{\mathbf{k}}-\mu _{AB} \\
\beta \gamma \sqrt{N}\varepsilon _{\mathbf{k}}-\mu _{AB} & \gamma^{2} \varepsilon _{\kk}
\end{array}
\right)
\end{equation}
The ``chemical potentials" $\mu=\mu_{A}=\mu _{B}$ and 
\begin{equation}
\mu _{AB}=\beta \gamma \sqrt{N}h_{AA}+\gamma ^{2}h_{AB}+\Omega \alpha \gamma
I_{A}
\end{equation}
is due to the constraint $\sum_{\mathbf{k}}\varphi _{A}(\mathbf{k})\varphi _{B}(\mathbf{k})=0$. The solutions $\varphi_{M=A,B}(\kk)$ to these equations must be the superposition of two functions $\eta_{\kk}/(e_{1,2}-\varepsilon_{\kk})$ that read
\begin{equation}
\varphi _{M}(\kk) = \frac{1}{\sqrt{V}} \sum_{\mu=1,2} \frac{c_{M,\mu}}{\sqrt{\mathcal{N}_{\mu\mu}}} \frac{\eta_{\kk}} {e_{\mu\mu}-\varepsilon_{\kk}}
\label{phs}
\end{equation}
where we have defined 
\begin{equation}
\mathcal{N}_{\mu\nu} = \int \frac{d^d\kk}{(2\pi)^{d}} \frac{\eta^{2}_{\kk}}{(e_{\mu}-\varepsilon_{\kk})(e_{\nu}-\varepsilon_{\kk})}
\end{equation}
We introduce two variables $t_{M}$ to connect the coefficients as $c_{M,2}=t_{M}c_{M,1}$. The constraints given by Eq.~(\ref{C2}) result in the relations
\begin{eqnarray}
c_{M,1} &=& \frac{1}{\sqrt{1+t^{2}_{M}+2\frac{\mathcal{N}_{12}}{\sqrt{\mathcal{N}_{11}\mathcal{N}_{22}}} t_{M}}} \nonumber \\
t_{B} &=& -\frac{\sqrt{\mathcal{N}_{11}\mathcal{N}_{22}}+t_{A}\mathcal{N}_{12}}{\sqrt{\mathcal{N}_{11}\mathcal{N}_{22}}t_{A}+\mathcal{N}_{12}}
\end{eqnarray}

To find optimized variational parameters, we construct the wavefunctions $\varphi _{M}(\kk)$ in Eq.~(\ref{phs}) and the matrix $\mathcal{E}_{0}$ in Eq.~(\ref{E0}) using different sets of $e_{1}$, $e_{2}$ and $t_A$. For a given set of $e_{1}$, $e_{2}$ and $t_A$, the ground state energy is given by $E_{\mathrm{GS}}=(N-1)h_{AA}+\lambda _{0}$, where $\lambda_0$ is the lowest eigenvalue of $\mathcal{E}_{0}$. By minimizing $E_{\rm GS}$, we obtain the optimal variational parameters.

\bibliography{Sci,books}
\bibliographystyle{naturemag}

\end{document}